# LOW-MASS QUARK STARS OR QUARK WHITE DWARFS

G. B. Alaverdyan*, A. R. Harutyunyan, and Yu. L. Vartanyan

*An equation of state is considered that, in superdense nuclear matter, results in a phase transition of the first kind from the nucleon state to the quark state with a transition parameter $\lambda > 3/2$ $\left(\lambda = \rho_Q / \left(\rho_N + P_0/c^2\right)\right)$. A calculation of the integrated parameters of superdense stars on the basis of this equation of state shows that on the stable branch of the dependence of stellar mass on central pressure $(dM/dP_c > 0)$, in the low-mass range, following the formation of a tooth-shaped break ($M = 0.08\ M_\odot$, $R = 200$ km) due to quark formation, a new local maximum with $M_{max} = 0.082\ M_\odot$ and $R = 1251$ km is also formed. The mass and radius of the quark core of such a star turn out to be $M_{core} = 0.005\ M_\odot$ and $R_{core} = 1.7$ km, respectively. Mass accretion in this model can result in two successive transitions to a neutron star with a quark core, with energy release like supernova outbursts.*

## 1. Introduction

Wittten's hypothesis that it is possible for strange quark matter to form in superdense nuclear matter [1] has stimulated numerous papers devoted both to the problem of the transition to such a state and the construction of a realistic equation of state and to the application of different versions of the equation of state for the study of compact objects containing strange quark matter [2-11]. It turned out that, depending on the values of the quantum chromodynamic (QCD) constants occurring in nature, two types of essentially different stellar configurations can exist: so-called strange stars, consisting entirely of strange quark matter, and hybrid stars, simulatneously containing neutron and strange quark matter. To study the functional dependence of the structural and integrated parameters of stellar configurations on the form of the equation of state of superdense matter, we considered an extensive set of realistic equations of state that provide for the coexistence of neutron matter with strange quark matter. In the low-mass range ($M/M_\odot \approx 0.08$) it was found that one of these equations of state results in the appearance of an additional local maximum on the curve of the dependence of the star's mass $M$ on its central pressure $P_c$, which makes possible the existence of a new family of stable, equilibrium stellar configurations with rather interesting distinguishing features. At the center of such a star there is a quark core, and the stellar radius can reach some 1000 km, which makes them resemble white dwarfs.

In this work we confine ourselves to a study of hybrid stars when just this equation of state is realized, concentrating our attention on the low-mass range. The results of a more detailed comparative analysis of the functional dependence of stellar properties on the choice of the equation of state in the entire possible range of variation of central pressure will be published later.

---

In considering the phase transition to quark matter, we assumed that an ordinary phase transition of the first kind occurs, and at some pressure $P_0$, corresponding to the coexistence of two phases, the energy density and baryon concentration undergo discontinuities. The mixed phase of quark and nuclear matter proposed by Glendenning [8] presumes, in contrast to the usual phase transition of the first kind, a continuous variation of pressure and density in the region of the appearance of the quark phase. If the mixed phase is energetically favorable, then instead of a phase transition of the first kind there are two phase transitions of the second kind, at the densities of the onset and end of the mixed phase, respectively.

Heiselberg et al. [9] showed that allowance for the local surface and Coulomb energies, which appear due to the formation of quark and nuclear structures in the mixed phase, makes it possible for the formation of a mixed phase to be energetically unfavorable. Thus, if the surface tension between the quark and nuclear matter is high enough, the appearance of a mixed phase is energetically unfavorable. The neutron star will then have a core of pure quark matter and a crust of nuclear matter. A phase transition of the first kind occurs in this case and the two phases coexist.

For low and moderate values of the surface tension of quark structures, it becomes possible for a mixed phase to exist, consisting of quark formations of various configurations in the form of drop-like, rod-like, and plate-like structures [10].

The uncertainty in the value of the surface tension of strange quark matter at present prevents a definite determination of which of these versions occurs in reality. Below we consider the case that presumes a surface tension that leads to a phase transition of the first kind with the possible coexistence of two phases.

## 2. Equation of State

The mass density inside a neutron star varies within a fairly wide range from several grams per cubic centimeter in the periphery (envelope) to $10^{15}$ g/cm$^3$ at the center. At present there is no unified theory that provides an adequate description of the state of such matter, with allowance for the formation of all possible constituents in the entire density range. In constructing an equation of state for the matter of a neutron star, therefore, one usually uses different equations of state for different density ranges, providing for continuity in the transition from one range to another, of course.

**2.1. Nucleon Matter (NM).** In the present work we have used the following equations of state for densities below the normal nuclear density:

$$7.86 \text{ g/cm}^3 < \rho < 1.15 \cdot 10^3 \text{ g/cm}^3 \quad \text{(FMT [12])},$$

$$1.15 \cdot 10^3 \text{ g/cm}^3 < r < 4.3 \cdot 10^{11} \text{ g/cm}^3 \quad \text{(BPS [13])}.$$

Starting with the density $\rho_{nd} = 4.3 \cdot 10^{11}$ g/cm$^3$, the composition of the matter changes due to neutron evaporation from neuclei, forming the so-called *Aen* structure, and the state is described by the equation

$$4.3 \cdot 10^{11} \text{ g/cm}^3 < r < 2.21 \cdot 10^{13} \text{ g/cm}^3 \quad \text{(BBP [14])}.$$

In the supernuclear density range we used the relativistic equation of state of neutron matter tabulated by Weber et al. [15], which was calculated with allowance for two-particle correlations on the basis of the Bonn meson-exchange potential [16]. That equation of state is designated as WGW–Bonn,

$$3.56 \cdot 10^{13} \text{ g/cm}^3 < \rho < 4.81 \cdot 10^{14} \text{ g/cm}^3 \quad \text{(WGW–Bonn [15])}.$$

Note that in joining this equation with the BBP equation of state, the overlapping density range $2.98 \cdot 10^{13}$ g/cm$^3 < \rho < 1.58 \cdot 10^{14}$ g/cm$^3$ was discarded from the latter.

These equations of state, which in combination cover the density range of $7.86$ g/cm$^3 < \rho < 4.81 \cdot 10^{14}$ g/cm$^3$, describes the matter of a neutron star with a nucleon structure.

To investigate the phase transition, we must know the dependence of the baryon chemical potential $\mu_B$ on pressure $P$, i.e., the function $\mu_B^{(NM)}(P)$, or the dependence of the baryon energy $\mathcal{E}$ on the baryon density $n$: $\mathcal{E}^{(NM)}(n)$. For this purpose, to the tabulated values of $P$, r, and $n$ we added values of the quantities

$$\mu_B^{(NM)}(P) = n\frac{\partial \rho c^2}{\partial n} - \rho c^2 = \frac{P + \rho c^2}{n}, \tag{1}$$

$$\mathcal{E}^{(NM)}(n) = \rho c^2 / n. \tag{2}$$

We used Aitken's interpolation scheme [17] to construct the continuous functions $\mu_B^{(NM)}(P)$, and $\mathcal{E}^{(NM)}(n)$.

**2.2. Strange Quark Matter (QM).** We used the quark bag model, developed at the Massachusetts Institute of Technology (MIT) [18], to describe the quark phase. The quark phase consists of three quark flavors, $u$, $d$, and $s$, and electrons in equilibrium with respect to weak interactions, provided by the reactions

$$d \rightarrow u + e^- + \bar{\nu}_e, \quad s \rightarrow u + e^- + \bar{\nu}_e,$$
$$u + e^- \rightarrow d + \nu_e, \quad u + e^- \rightarrow s + \nu_e. \tag{3}$$

These interactions lead to the following relations among the chemical potentials:

$$\mu_s = \mu_d = \mu_u + \mu_e. \tag{4}$$

The condition of electrical neutrality for the quark–electron plasma has the form

$$\frac{2}{3}n_u - \frac{1}{3}n_d - \frac{1}{3}n_s - n_e = 0. \tag{5}$$

The number density of particles of the $i$th kind is defined by the equation

$$n_i(\mu_i) = -\frac{\partial \Omega_i}{\partial \mu_i} \quad (i = u, d, s, e), \tag{6}$$

where $\Omega_i$ is the thermodynamic potential of particles of the $i$th kind. We used the expression for the thermodynamic potential $\Omega_i$ given in [2, 4] in a linear approximation with respect to the quark–gluon interaction constant $\alpha_c = g^2/4\pi$, where $g$ is the QCD binding constant. In [2, 4] the masses of $u$ and $d$ quarks are taken as zero while the mass of the strange quark is $m_s = 175$ MeV.

Equations (4)-(6) enable us to express the particle concentrations $n_i$ and thermodynamic potentials $\Omega_i$ as functions of the chemical potential of one of the quark flavors, of $\mu_s = \mu_d = \mu$, let us say.

In the MIT bag model, pressure is defined by the expression

$$P(\mu) = -\sum_{i=u,d,s,e} \Omega_i(\mu) - B, \tag{7}$$

where $B$ is the "bag" constant, characterizing the vacuum pressure and providing for confinement.

The following phenomenological parameters of the "bag" model are used in the present work:

$$m_s = 175 \text{ MeV}, \quad B = 55 \text{ MeV/fm}^3, \quad \text{and} \quad \alpha_c = 0.5.$$

The energy density $\rho^{(QM)}c^2$ of the quark–electron plasma and the baryon concentration $n_B^{(QM)}$ are defined by the expressions

$$\rho^{(QM)}c^2 = \sum_{i=u,d,s,e}(\Omega_i + \mu_i n_i) + B, \tag{8}$$

$$n_B^{(QM)} = \frac{1}{3}(n_u + n_d + n_s). \tag{9}$$

The baryon chemical potential $\mu_B^{(QM)}$ and the energy $\mathcal{E}^{(QM)}$ per baryon for strange quark matter are defined just as in the case of neutron matter:

$$\mu_B^{(QM)} = \frac{P + \rho^{(QM)} c^2}{n_B^{(QM)}}, \tag{10}$$

$$\mathcal{E}^{(QM)} = \rho^{(QM)} c^2 / n_B^{(QM)}. \tag{11}$$

**2.3. Phase Transition of the First Kind.** A phase transition of the first kind occurs in the model that we are considering. The Gibbs conditions

$$\begin{aligned} P^{(NM)} &= P^{(QM)} = P_0, \\ \mu_B^{(NM)} &= \mu_B^{(QM)} \end{aligned} \tag{12}$$

enable us to find the pressure $P_0$, the baryon number densities $n_N$ and $n_Q$, and the mass densities $\rho_N$ and $\rho_Q$ characterizing the two coexisting phases.

The parameters of a phase transition of the first kind can also be determined from the standard Maxwellian construct. The functional dependence of the energy per baryon in a phase transition of the first kind satisfies the analogous Gibbs conditions through the equations

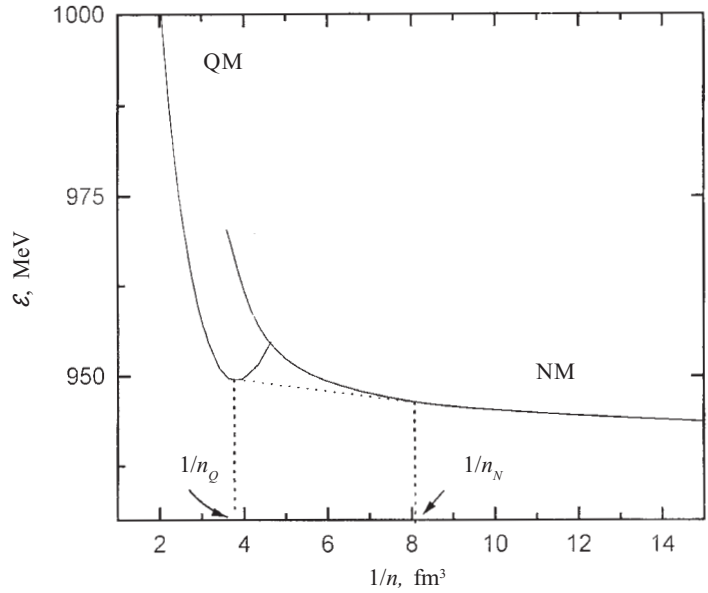

Fig. 1. Maxwellian construct for a phase transition of neutron matter (NM) to strange quark matter (QM). $\mathcal{E}$ is the energy per baryon and $n$ is the baryon density.

$$\frac{\partial \mathcal{E}^{(NM)}}{\partial \left(1/n_B^{(NM)}\right)} = \frac{\partial \mathcal{E}^{(QM)}}{\partial \left(1/n_B^{(QM)}\right)} = -P_0, \tag{13}$$

$$\mathcal{E}_Q - \mathcal{E}_N = P_0 \left(\frac{1}{n_N} - \frac{1}{n_Q}\right), \tag{14}$$

which corresponds to the common tangent to the graph of the energy $\mathcal{E}$ per baryon as a function of $1/n$. The Maxwellian construct enabling one to determine the parameters $P_0$, $n_N$, and $n_Q$ is given in Fig. 1.

Numerical calculations within the framework of this model led to the following values of the characteristics of a phase transition of the first kind: $P_0$ = 0.76 MeV/fm³, $n_N$ = 0.12 fm⁻³, $n_Q$ = 0.26 fm⁻³, $\rho_N c^2$ = 113.8 MeV/fm³, $\rho_Q c^2$ = 250.5 MeV/fm³.

## 3. Models of Neutron Stars with a Core of Strange Quark Matter (Hybrid Stars)

**3.1. TOV and Hartle Equations.** The space-time geometry for spherically symmetric, static stars is given by the metric

$$ds^2 = e^{\nu(r)} c^2 dt^2 - e^{\lambda(r)} dr^2 - r^2 \left(d\vartheta^2 + \sin^2\vartheta\, d\varphi^2\right). \tag{15}$$

The structure function $e^{\lambda(r)}$ is defined by the equation

$$e^{\lambda(r)} = \left(1 - 2Gm(r)/(c^2 r)\right)^{-1},$$

where $m(r)$ is the mass accumulated inside a sphere of radius $r$. The metric function $\nu(r)$ outside the star satisfies the condition

$$e^{\nu(r)} = 1 - 2GM/(c^2 r), \qquad r > R. \tag{16}$$

The Einstein equations for a static star lead to the well-known Tolman-Oppenheimer-Volkoff (TOV) equations [19-22]

$$\frac{dP}{dr} = -\frac{G(\rho + P/c^2)}{r^2 \left(1 - 2Gm(r)/(c^2 r)\right)} \left(m + 4\pi r^3 P/c^2\right), \tag{17}$$

$$\frac{dm}{dr} = 4\pi r^2 \rho, \tag{18}$$

$$\frac{dm_0}{dr} = \frac{4\pi r^2 \rho_0}{\sqrt{1 - 2Gm(r)/(c^2 r)}}, \tag{19}$$

$$\frac{dm_p}{dr} = \frac{4\pi r^2 \rho}{\sqrt{1 - 2Gm(r)/(c^2 r)}}, \tag{20}$$

$$\frac{d\nu^*}{dr} = \frac{2G\left(m + 4\pi r^3 P/c^2\right)}{c^2 r^2 \left(1 - 2Gm(r)/(c^2 r)\right)}. \tag{21}$$

Here $m$ and $m_p$ are the gravitational and proper masses, respectively, while $m_0$ is the rest mass within a sphere of radius $r$, $\rho_0 = \dfrac{M(^{56}Fe)}{56} n$ is the rest-mass density, and $n$ is the baryon number density.

It is convenient to replace the equation for the moment of momentum, which is a second-order differential equation [23], by a system of two first-order equations* [24],

$$\frac{d\omega^*}{dr} = \frac{6 G l(r) e^{\nu^*/2}}{c^2 r^4 \sqrt{1 - 2 G m(r)/(c^2 r)}}, \tag{22}$$

$$\frac{dl}{dr} = \frac{8}{3} \frac{\pi \omega^*(r) r^4 (\rho + P/c^2) e^{-\nu^*/2}}{\sqrt{1 - 2 G m(r)/(c^2 r)}}. \tag{23}$$

Integration begins from the center of the star with the boundary conditions $P = P_c$, $m(0) = 0$, $m_0(0) = 0$, $m_p(0) = 0$, $\nu^*(0) = 0$, $\omega^*(0) = \omega_0$, and $l(0) = 0$, where $\omega_0$ is an arbitrary constant.

The boundary of the equilibrium configuration is determined by the condition $P(R) = 0$, where $R$ is the star's coordinate radius. The total gravitational mass $M$, total rest mass $M_0$, and total proper mass $M_p$ are defined to be $M = m(R)$, $M_0 = m_0(R)$, and $M_p = m_p(R)$.

The structure function $\nu(r)$ is defined by the expression

$$\nu(r) = \nu^*(r) - \nu^*(R) + \ln(1 - 2 G m(r)/(c^2 r)). \tag{24}$$

The functions $\omega^*(r)$ and $l(r)$ enable one to determine the star's moment of inertia:

$$I = \frac{l(R)}{\omega^*(R)\sqrt{1 - 2 G M/(c^2 R)} e^{-\nu^*(R)/2} + \dfrac{2 G l(R)}{c^2 R^3}}. \tag{25}$$

The integrated parameters of a superdense star are determined by numerical integration of the system of equations (17)-(23) for a given equation of state $\rho(P)$ and $\rho_0(P)$.

**3.2. Results of Numerical Integration.** We integrated the TOV and Hartle equations, using the equation of state of superdense matter described above, with allowance for the possible appearance of strange quark matter as a result of a phase transition of the first kind.

The results of a calculation of the total mass $M$, rest mass $M_0$, and proper mass $M_p$ as functions of central pressure $P_c$ are given in Fig. 2. Whereas in the range of maximum mass (configuration *f*) these curves have the usual character, in the low-mass range, where stability loss again occurs — the condition $dM/dP_c > 0$ (configuration *a*) is violated — the curve has a number of peculiarities that are absent in the case of other equations of state. This range is given on an enlarged scale in the upper left corner of the figure. Immediately after configuration *a* there is a tooth-shaped break on the curve (configuration *b*), due to quark formation. The section *ab* corresponds to stable neutron stars with no quark core. Configurations with small quark cores are unstable (the section *bc* of the curve, where we have $dM/dP_c < 0$). This agrees with the results of Kaempfer [25], who showed that if the condition

$$\lambda = \frac{\rho_Q}{\rho_N + P_0/c^2} > 3/2 \tag{26}$$

is satisfied, configurations with a low-mass core of a new phase are unstable. In the case under consideration, we have $\lambda = 2.19$, i.e., 1 is in accord with the condition (26).

---

* A factor of 6 was omitted from Eq. (22) in [24].

Usually, when the condition (26) is satisfied, the tooth-shaped break *abc* occurs not in the low-mass range but on the ascending branch of the $M(P_c)$ curve, and the curve has a monotonically ascending character after configuration *c* up to the maximum-mass configuration *f*. But in the case under consideration, immediately after this break, again in the low-mass range, a local maximum is formed, configuration *d*, the radius of which exceeds 1000 km, while its mass, although slight, exceeds the mass of configuration *b* and is 0.082 $M_\odot$. Along with the radius of the identified maximum for this configuration, there is also a moment of inertia (see Table 1 and Fig. 6).

In Table 1 we give the main parameters of the critical configurations *a*, *b*, *c*, *d*, *e*, and *f*. There we also give the packing factor a. As we see, this quantity has a positive sign for all the critical configurations and, except for configuration *f*, has the same order of magnitude as for white dwarfs.

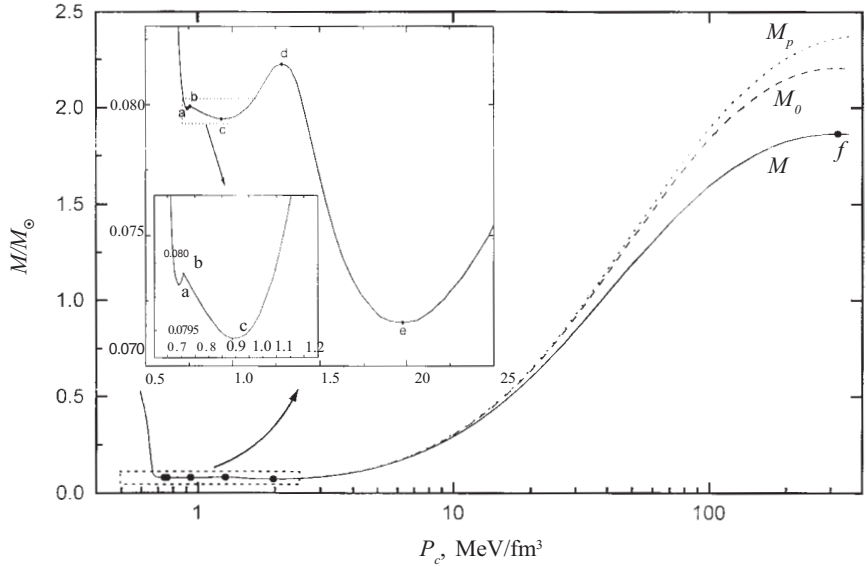

Fig. 2. Total mass $M$, rest mass $M_0$, and proper mass $M_p$ as functions of central pressure. The behavior of $M(P_c)$ in the low-mass range is shown on an enlarged scale in the upper left-hand corner. Critical configurations are denoted by the symbols *a*, *b*, *c*, *d*, *e*, and *f*; *a* corresponds to an ordinary neutron star with the minimum mass and *b* to the threshold for the formation of a quark core.

TABLE 1. Parameters of Critical Configurations

| Configurations | $P_c$, MeV/fm$^3$ | $\dfrac{M}{M_\odot}$ | $R$, km | $\alpha = \dfrac{M_0 - M}{M_0}$ | $I$, $M_\odot \cdot$km$^2$ | $\dfrac{M_{core}}{M_\odot}$ | $R_{core}$, km |
|---|---|---|---|---|---|---|---|
| *a* | 0.74 | 0.0798 | 254.7 | 0.00573 | 9.99 | 0 | 0 |
| *b* | 0.76 | 0.080 | 205 | 0.00597 | 6.6 | 0 | 0 |
| *c* | 0.94 | 0.079 | 380 | 0.00576 | 25.4 | 0.001 | 1.0 |
| *d* | 1.3 | 0.082 | 1251 | 0.00622 | 861.4 | 0.005 | 1.73 |
| *e* | 1.97 | 0.072 | 133.2 | 0.00596 | 2.4 | 0.016 | 2.59 |
| *f* | 321 | 1.86 | 10.8 | 0.15495 | 94.1 | 1.85 | 10.26 |

The mass density r as a function of the *r* coordinate for configuration *d* is given in Fig. 3. The dashed line shows the boundary of the strange quark core and the dotted line corresponds to the threshold for neutron evaporation from nuclei (the limit of the *Aen* plasma). There is a density jump at the boundary of the quark core ($\rho_Q c^2 = 250.5$ MeV/fm$^3$, $\rho_N c^2 = 113.8$ MeV/fm$^3$).

As follows from the calculations, the threshold for the formation of an *Aen* plasma corresponds to a radial coordinate $R_{nd} = 13.24$ km and an accumulated mass $M_{nd} = 0.07\, M_\odot$. Note that this configuration is similar in size to a white dwarf, but most of its mass is concentrated in the *Aen* plasma.

The dependence of a star's mass *M* on its radius *R* is shown in Fig. 4. The symbols *a*, *b*, *c*, *d*, *e*, and *f* denote the same configurations as in Fig. 2. It is seen from Fig. 4 that stars of the same mass that correspond to the two branches

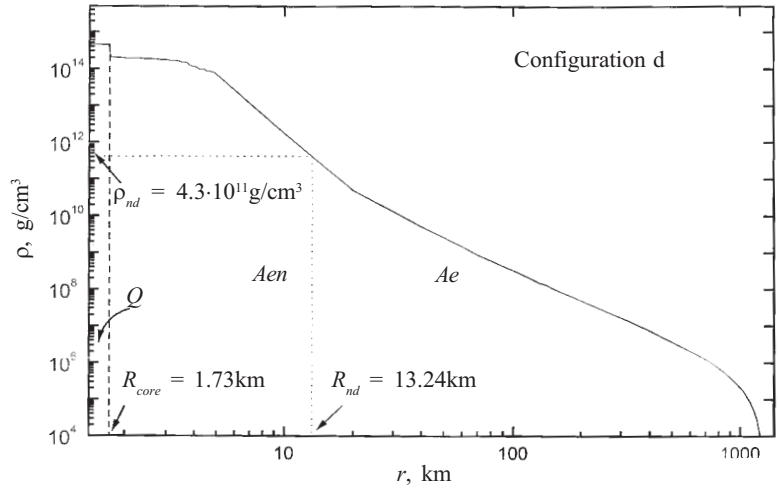

Fig. 3. Mass density r as a function of the radial coordinate *r* for configuration *d* (see Fig. 2 and Table 1). The dashed line corresponds to the boundary of the strange quark core and the dotted line corresponds to the threshold for the formation of an *Aen* plasma at the density $\rho_{nd} = 4.3 \cdot 10^{11}$ g/cm$^3$.

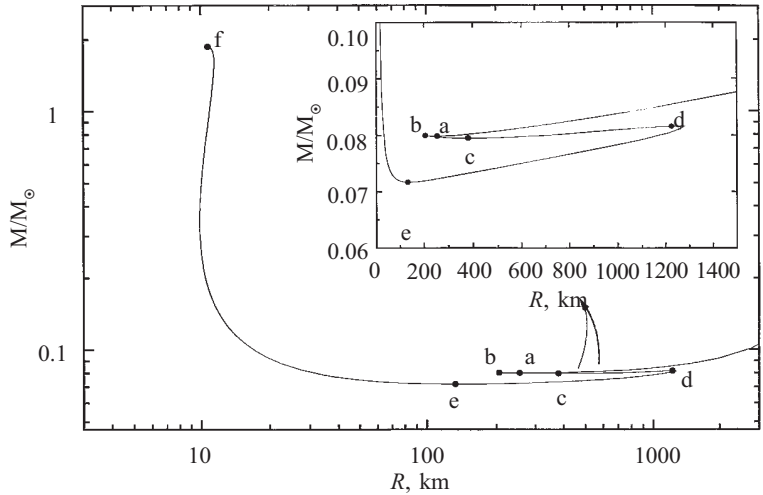

Fig. 4. Dependence of a star's mass *M* on its radius *R*. The behavior of *M(R)* in the low-mass range is shown on an enlarged scale in the upper right-hand corner. *a*, *b*, *c*, *d*, *e*, and *f*: same configurations as in Fig. 2.

*cd* and *ef* differ fairly strongly from one another in radius. Whereas stars on the *ef* branch have radii of ~10 km, stars on the *cd* branch have fairly large radii of some 1000 km, typical of white dwarfs.

In Fig. 5 we show the stellar radius $R$ and the radius $R_{core}$ of the quark core as functions of the central pressure $P_c$. The curve corresponding to neutron stars having no quark core is shown by a dotted line. The radius of the quark core as a function of the central pressure is shown by a dashed line. The dash–dot line shows the distance from the star's center to the threshold point where an *Aen* plasma is formed at a density $\rho_{nd} = 4.3 \cdot 10^{11}$ g/cm$^3$ as a result of neutron evaporation from nuclei. It is seen that a clearly defined maximum is observed in the region of configuration *d*.

A graph of the relativistic moment of inertia $I$ as a function of $P_c$ is given in Fig. 6.

It must be noted that if the equation of state under consideration is realized, mass accretion onto a neutron star will result in two successive discontinuous transitions to a neutron star with a quark core, as a result of which two successive processes of energy release will occur. A star with a quark core, belonging to the *cd* branch, is formed first; further accretion results in configurations with a radius on the order of 1000 km, and finally, as a result of a second catastrophic reorganization, a star on the *ef* branch is formed, having a radius of some 100 km.

We note in conclusion that to clarify the regularity of our result and that it is no accident that an additional maximum (although a barely distinguishable one) appears on the $M(P_c)$ curve, we considered several trial equations of state of neutron matter that differ from the analyzed equation in the region near the threshold for formation of the quark phase. The investigations confirmed the regularity of the result and showed that varying the equation of state in the range of $9 \cdot 10^{13}$ g/cm$^3 < \rho < 1.8 \cdot 10^{14}$ g/cm$^3$ can even result in enhancement of the detected feature on the $M(P_c)$ curve in some cases.

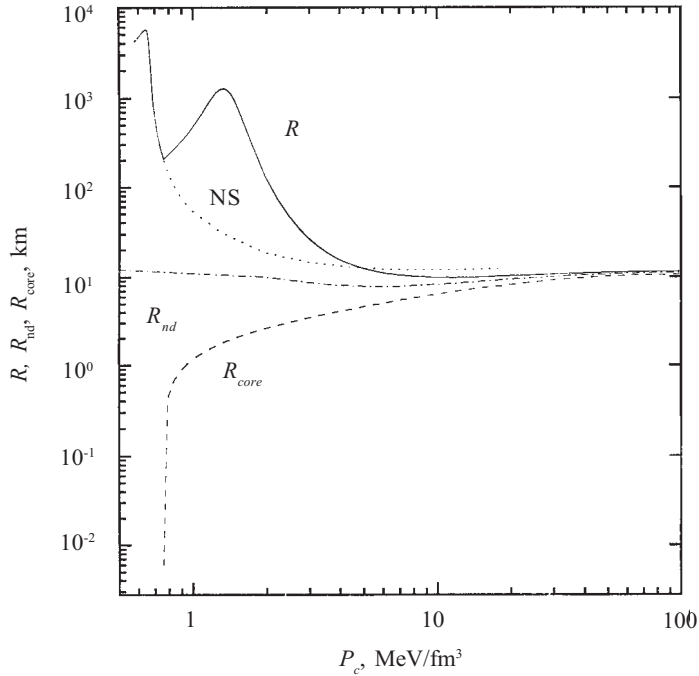

Fig. 5. Stellar radius $R$ as a function of central pressure $P_c$. The dotted line corresponds to ordinary neutron stars having no quark core. The dashed line shows the dependence of the radius $R_{core}$ of the quark core on $P_c$ while the dash–dot line shows the dependence of the coordinate $R_{nd}$ corresponding to the threshold for the formation of an *Aen* plasma.

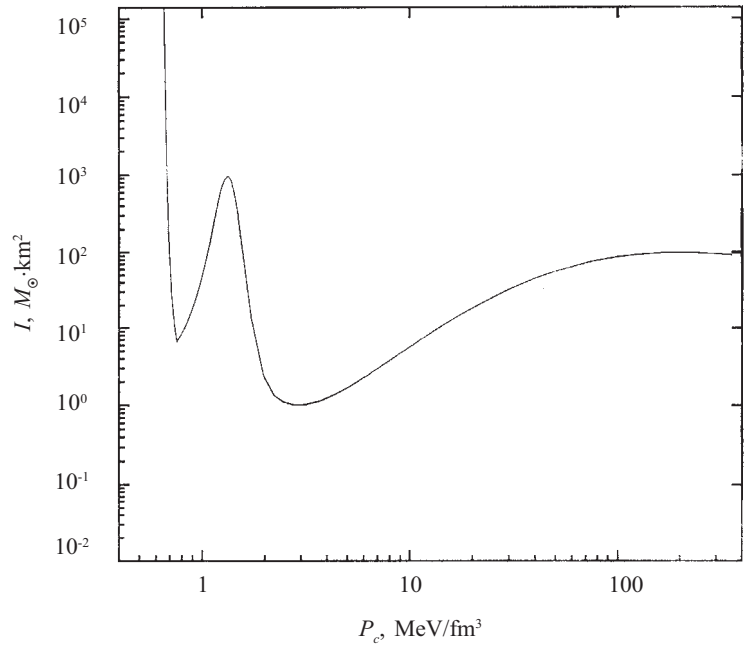

Fig. 6. Relativistic moment of inertia $I$ as a function of central pressure.

## 4. Conclusion

A phase transition of the first kind in superdense nuclear matter, from the nucleon component to the strange quark state, with a transition parameter $\lambda > 3/2$ usually results in the appearance of a small tooth-shaped break on the stable branch of the dependence of stellar mass on central pressure [the $M(P_c)$ curve]. In the model considered above, in which stability loss in the low-mass range (violation of the condition $dM/dP_c > 0$) occurs at relatively higher densities than in other models ($\rho_c = 2\cdot10^{14}$ g/cm$^3$, Table 1, configuration *a*) and is next to the threshold for quark creation ($\rho_c = 4.5\cdot10^{14}$ g/cm$^3$, Table 1, configuration *b*), a new local maximum arises, resulting in the possible existence of low-mass, superdense stars with a radius exceeding 1000 km and having a 1-km quark core in which only 6% of the entire stellar mass is concentrated. Such stars are similar in size to white dwarfs, while most of their mass is concentrated in the *Aen* phase.

This work was carried out under theme N2000-55, supported by the Ministry of Education and Science of the Republic of Armenia.